\documentclass[final,lefttitle,10pt]{elsarticle}

\usepackage[utf8]{inputenc}
\usepackage[T1]{fontenc}
\usepackage{graphicx}
\usepackage{subfigure}
\usepackage{fullpage}
\usepackage{hyperref}
\usepackage{xspace}
\usepackage{microtype}
\usepackage{soul}
\usepackage{xspace}

\usepackage{algorithmic}
\usepackage{algorithm}
\floatname{algorithm}{Listing}

\newcommand{\docName}{report\xspace}

\usepackage[subfigure]{tocloft}

\newcommand{\listtodoname}{List of TODOs}
\newlistof{todo}{todos}{\listtodoname}

\newcommand{\secref}[1]{Section~\ref{#1}\xspace}
\newcommand{\lstref}[1]{Listing~\ref{#1}\xspace}

\begin{document}

\journal{arXiv}
\begin{frontmatter}

\title{Cooperation in public goods game on square lattices with agents changing interaction groups}
\author{Jaros{\l}aw Adam Miszczak}
\address{Institute of Theoretical and Applied Informatics, Polish Academy
of Sciences, Ba{\l}tycka 5, 44-100 Gliwice, Poland}
\ead{jmiszczak@iitis.pl}
\ead[orcid]{https://orcid.org/0000-0001-8790-101X}

\begin{abstract}
The emergence of cooperation in the groups of interacting agents is one of the most fascinating phenomena observed in many complex systems studied in social science and ecology, even in the situations where one would expect the agent to use a free-rider policy. This is especially surprising in the situation where no external mechanisms based on reputation or punishment are present. One of the possible explanations of this effect is the inhomogeneity of the various aspects of interactions, which can be used to clarify the seemingly paradoxical behaviour. In this work we demonstrate that the diversity of interaction networks helps to some degree explaining the emergence of cooperation. We extend the model of spatial interaction diversity by enabling the evaluation of the interaction groups. We show that the process of the reevaluation of the interaction group facilitates the emergence of cooperation. Furthermore, we also observe that a significant participation of agents switching their interaction neighbourhoods has a negative impact on the formation of cooperation. The introduced scenario can help to understand the formation of cooperation in the systems where no additional mechanisms for controlling agents are included.

\end{abstract}

\end{frontmatter}


\section{Introduction}

Cooperation is one of the emergent phenomena puzzling researchers in many areas of science and the question about mechanisms supporting the emergence of cooperation is among the central questions of modern theories describing social, economical, and biological  systems~\cite{axelrod1981evolution,bergstrom2002evolution,nowak2006five,ostrom2015governing}.  One of the paradigms used to tackle the questions how costly, cooperative behaviours can appear in complex systems is based on evolutionary game theory \cite{smith1986evolutionary,szabo2007evolutionary,sandholm2017evolutionary,newton2018evolutionary}. Standard models used to study the emergence of cooperation include Prisoner's Dilemma (PD), Snowdrift Game, and Public Goods Game (PGG) \cite{sigmund2010calculus}. In this approach one aims at combining the decisions made by individual agents to provide a dynamic perspective on the cooperation. This is used in connection with the utilization of mathematical apparatus developed in physical sciences to successfully understand the effects observed in economics \cite{farmer2005is} and collective behaviour in human populations~\cite{galam1982sociophysics, schweitzer2018sociophysics}.

However, the models based on the evolutionary game theory cannot be treated exactly in situations that are realistic for evolving populations~\cite{adami2016evolutionary}, and agent-based methods are becoming an important tool for simulating the complex behaviours in heterogeneous populations~\cite{tisue2004netlogo,farmer2009economy,edmonds2013simulating,daly2022quo}.


One of the seminal works considering the inhomogeneity of interactions was dealing with Prisoner's Dilemma~\cite{perc2008social}. In this case, the authors demonstrated that social diversity, introduced in the form of the scaled payoff, promotes cooperation in the case of spatial Prisoner's Dilemma game. Hence, it is natural to consider the inhomogeneity of spatial interaction as one of the important factors in the mechanics of cooperation. In the case of Prisoner's Dilemma this approach has been further developed to include factors such as indirect interactions \cite{hu2015incorporating} or memory~\cite{liu2010memory, wang2016cooperation, takahara2023twisted}. 
Random and dynamic groupings were also studied in the context of multiplayer snowdrift games~\cite{xu2022enhanced}, where the dynamics behaviour was found to enhance the cooperation between agents.

Particularly important example of the agent-based modelling used to study social behaviour is given by the Public Goods Game. This model displays many features observed in real systems of interacting agents, and many attempts to extend it with mechanisms explaining cooperation have been proposed, including several attempts based on the diversification of population.

The effect of spatial interaction diversity in Public Goods Game was analysed in
\cite{santos2008social}. In this case, social diversity was introduced by the
means of heterogeneous graphs, and the  special role of scale-free networks was
highlighted~\cite{santos2005scale}. In \cite{zhang2025zero} a method for designing  zero-determinant strategy in multiplayer two-strategy repeated games under implementation errors based has been proposed. In \cite{zhang2025co} co-evolutionary dynamics based on threshold Public Goods Games under collective-risk has been considered for exploring collective cooperation in the context of collective-risk dilemmas.

Recently, the behaviour of the Public Goods Games in the model enabling the migration between the interaction groups was studied in~\cite{tomassini2021computational}. The results obtained in \cite{tomassini2021computational} suggest that by enabling the possibility of reevaluating the interaction neighbourhood one can support the formation of cooperation and help avoiding free-riding.  In the recent work \cite{ma2023evolution}, it was also observed that groups interact in heterogeneous environments can be more cooperative. Another example was provided in \cite{bittencourt2023interplay} where a simple model of individual careers based on the concept of random walk on a hexagonal lattice has been proposed. This provides a strong argument for connecting dynamical behaviour of agents and with their approach to cooperation. 

In the model of the Public Goods Game on lattices studied in \cite{hauert2002replicator} agents can randomly select iteration partners in the von Neumann or Moore neighbourhood. This effect has been shown to contribute to the cooperation level when compared with the traditional model in which the social scope is homogeneous. More recently, a similar effect has been observed in spatially inhomogeneous variant of the spatial Public Goods Game was considered in~\cite{shang2022cooperation}. In this case, the spatial interaction diversity was based on the diversity in the selection of the neighbourhoods. This mechanism also had a positive impact on the emergence of cooperation. Hence, one can argue that the spatial inhomogeneity, defined as the ability of altering the interaction group, can indeed contribute to the promotion of cooperative behaviour.

However, one should also note that some aspects of diversity can be potentially disadvantageous for the cooperation in the Public Goods Game. This was studied in the heterogeneous model~\cite{flores2023heterogeneous}, where the diversity in the contribution leads to the increase of the synergy factor value required to form cooperation. Hence, one can see that the mobility of agents can also have a negative impact on the cooperation.

Considering the above, one can see that the diversity in the interactions can be used to explain the emergence of cooperation even in the situation where the mechanisms based on individual reputation and punishment are not present. At the same time, the impact of this aspect is far from being fully understood. Following this line of research, in this \docName we study the scenario of spatial interaction diversity in the case of Public Goods Game. To this end, we introduced the model based on the ones studied in \cite{tomassini2021computational} and \cite{shang2022cooperation}, and we stress two important elements of the dynamics in the interaction neighbourhood selection. First, we introduce a model where each agent is assigned a different size of the interaction group. This enables us to study the impact of the size of the interaction group and the promotion of cooperation. Next, we consider the existence of a subpopulation of roaming agents, which can re-evaluate their choice of the interaction group. Hence, we study the dynamics of the spatial interaction diversity where the subpopulation of agents has the ability to alter the interaction neighbourhood. Such mode can be illustrated by the behaviour of user of the social network who have the ability and tendency to alter their connections dynamically or by the presence of generalist pollinator in pollination network.

This \docName is organized as follows. In \secref{sec:preliminaries} we introduce notation and core elements of the considered models and provide some literature relevant for the concepts studied in this work. In \secref{sec:model} we describe a model of public goods game with spatial interaction diversity. Next, in \secref{sec:results} we present a series of numerical experiments demonstrating the interplay between the dynamics of spatial behaviour and the emergence of cooperation. Finally, in \secref{sec:final} we summarize the presented results and discuss the relation between the introduced approach and the previously proposed models.

\section{Preliminaries}\label{sec:preliminaries}

Public Goods Game (PGG) is one of the prominent models used to study the interaction between the agents in the situation of common pool resources~\cite{axelrod1981evolution, ostrom2015governing}. 
In the original PGG, each of $N$ players participating in the process has an opportunity to decide whether to contribute to a common pool or not. Each player can contribute a fixed amount of resources, $c$, to the common pool. The players deciding to do so are called contributors. However, some players might decide not to contribute and choose a free-rider strategy. The total contribution is multiplied by a synergy factor $r$, and equally divided between all players. It is easy to see that becoming a free-rider is the Nash equilibrium because any player will do better by choosing it, regardless of the choices made by other players. This leads to the social dilemma.

Let us consider a basic version of the Public Goods Game (PGG) with $N$ players on square $L\times L$ lattice with periodic boundary conditions. Initially, each player is assigned to the group of cooperators, with strategy $s=C$, or defectors, with strategy $s=D$, and a neighbourhood $G$. In the standard version of the game, each player is assigned the same neighbourhood, usually von Neumann neighbourhood or Moor neighbourhood.

For each elementary game, a game is played between a player and the players from his neighbourhood. Each player can contribute a fixed amount of resources, $c$, to the pool. During the game, the player and his neighbours contribute an amount $c=1$, for each one with strategy $C$, or $c=0$ for each one with strategy $D$. Players deciding to do so became \emph{contributors}, and players who decide not to contribute and become \emph{free-riders}. One should note that the variant of the scenario with heterogeneous contributions have also been considered~\cite{flores2023heterogeneous}.

Next, the total contribution is multiplied by a synergy factor $r$, and equally divided between all players. Thus, for the player $k$, the payoff from the elementary game with all his neighbours reads
\begin{equation}
\Theta_{k} = \sum_{i\in G_k}\frac{r \left(\sum_{j\in G_i} c_{j}  + c_{i} \right) - c_k}{|G_i|+1}
\end{equation}

The probability of player $i$ to adopt the strategy of player $j$ is given by the Fermi distribution
\begin{equation}
	p_{F}(s_i\leftarrow s_j) = \frac{1}{1+\exp{\left(-\frac{\Theta_j-\Theta_i}{\kappa}\right)}},\label{eqn:imitation-fermi}
\end{equation}
where $\Theta_i$ and $\Theta_j$ are payoffs of player $i$ and player $j$, respectively, and $\kappa$ the noise. For the case $\kappa\rightarrow0$, player $i$ will copy the strategy iff $\Theta_j>\Theta_i$, and for $\kappa\rightarrow\infty$, the strategy is copied using a coin toss. The detailed analysis of the role of noise on the dynamics of Public Goods Game can be found in~\cite{vukov2006cooperation, szabo2009selection, javarone2016role}. We also consider variants with the synchronous and asynchronous strategy update ~\cite{wang2022local}, which can have a significant impact on  the stability of the system~\cite{miszczak2023rule}.

Another function which can be used to control the imitation process is given by the \emph{difference-of-payoffs} formula
\begin{equation}
p_\Delta(s_i\leftarrow s_j) = 
\left\{
\begin{array}{lc}
0 & \Theta_i > \Theta_j\\
\frac{1}{2} & \Theta_i = \Theta_j \\
1 & \Theta_i < \Theta_j \\
\end{array}
\right..\label{eqn:imitation-diff}
\end{equation}
The discussion of the rationale behind using the imitation function can be found in \cite{jusup2022social} and the arguments supporting the particular functions can be found in~\cite{traulsen2010human} and \cite{grujic2020do}. In this \docName we focus mostly on the imitation process governed by the Fermi function, and difference-of-payoffs is used only to demonstrate some of the observations.

Clearly the above scenario is oversimplified, and cannot be expected to provide a solid model for the cases where the population consists of inhomogeneous agents, having different preferences concerning social interactions and characterized by various individual abilities. Hence, several methods for introducing diversity in the PGG have been proposed, including reputation~\cite{shen2022high-reputation,bi2023heterogeneous}, mobility \cite{sicardi2009random,valverde2017global},  differentiation in the imitation ability~\cite{amaral2018heterogeneous},  heterogeneous contributions of the agents~\cite{flores2023heterogeneous}, introduction of selective agents \cite{huang2023evolution}, and varying sizes of interaction neighbourhoods~\cite{xu2023evolution}.

The focus of this \docName is on the social diversity understood as the diversity in the interaction groups~\cite{santos2005scale,sicardi2009random,shang2022cooperation}. The results obtained in this \docName are focused on the spatial interaction inhomogeneity, with each agent assigned an individual groups of agents to interact with. Additionally, the spatial interactions of the agents are diversified by introducing the reevaluation in the selection of the interaction neighbourhoods. 

For the purpose of the presented considerations we introduced the following concepts. 

\emph{Spatial neighbourhood} of the agent is the set of agents in the proximity of the agent. In the local scenario of choosing the spatial neighbourhood agents are limited to their nearest neighbours, while in the global case the are no limitation on the spatial neighbourhood. Hence, local scenario corresponds to the geometry of 2D grid, and the global case corresponds to the complete graph of potential interactions. 

\emph{Interaction group} of the agent is the set of agents belonging to the spatial neighbourhood of the agent, which are participating in the process governing the evolutionary dynamics. One can consider a scenario where the interaction group is equal to the spatial neighbourhood. However, in the presented \docName, the interaction group can be chosen not only from the spatial neighbourhood. Indeed, we are mostly interested in the case where the interaction group can be different for each agent participating in the process. Additionally, we also consider the case where interaction group is time-dependent and it can be altered during the evolution.

\emph{Roaming agent} is the agent which is assigned a probability of choosing a new interaction group before engaging in the game. In the presented \docName we limit our considerations to the case of a fixed probability of switching the interaction group assigned to all roaming agents. One should also note that, in the current context, the roaming agent does not have to change the spatial neighbourhood or display any kind of mobility. As we focus on the interactions with other agents, the roaming is understood as the possibility to alter the interaction group.

Also, as it was already pointed in the introduction, there are many aspects of diversity which can be taken into account in models involving public goods game. In this \docName, however, we will only deal with the diversity in the network of interactions. Hence the term diversity will be always related to the changes in the connection between the agents, and the connections represent the interactions during the elementary games.

\section{Model of spatial interaction diversity}\label{sec:model}

As it has been already mentioned in the introduction, some of the authors suggested that the interaction diversity is one of the factors leading to the promotion of cooperation in various scenarios involving evolutionary games on graphs. This observation was also confirmed in  the case of PGG. In particular, in~\cite{santos2008social} it was demonstrated that the diversity of the neighbourhood selection can promote the cooperation among the agents. In this model the diversity was introduced by the means of heterogeneous graph. In this case the cooperation was promoted by the increasing diversity in the size of the interaction neighbourhoods.

The model of spatial interaction diversity based on the square lattice was proposed in~\cite{shang2022cooperation}, where it was demonstrated that the by diversifying the interaction neighbourhoods one is able to decrease the synergy factor necessary for the cooperative strategy to become dominant.  The model used to show this in~\cite{shang2022cooperation} was based on the assumption that each participant can choose its own groups of agents to interact with. 

In this \docName we provide an extension of the model proposed in~\cite{shang2022cooperation} by introducing the mechanisms for incorporating into the model the dynamics of the selection of the interaction neighbourhoods. To this end, we include two generalizations of this model developed in ~\cite{shang2022cooperation} which enable us to take into account the diversity in the spatial interactions within the population.

In the first extension, in order to take into account the local dynamics of interactions, we introduce the concept of \emph{roaming agents}. Roaming agents are reevaluating their interaction neighbourhood during the evolution. We study the impact of the participation of the roaming agents in the population on the
cooperation. Thus, we generalize the spatial diversity model introduced in~\cite{shang2022cooperation} with the local dynamics of the neighbourhood selection.

In the second extension we focus on the global diversity in the spatial interaction. To achieve this and to make the model independent on the particular graph structure, we propose a generalized version of the spatial interaction diversity where the interaction neighbourhoods are selected from the full population. In this case we also study the impact on roaming agents on the formation of cooperative behaviours. Hence, this extension follows the lead of \cite{santos2008social} by enabling us to relate the diversity in the network structure to the promotion of cooperation.

Such a model can be framed in the context of social networks, where agents are users and the connections are friendships, follows, or group memberships. Imagine a professional networking platform such as LinkedIn. Each user is an agent connected to others through their contacts. Most users have relatively stable connections: once connected, they rarely drop or change contacts drastically. Such agents connect with colleagues and friends, and their network stays fairly stable. A roaming agent, however, has the ability and tendency to alter their connections dynamically. He actively seeks out new groups, disconnects from old ones, and forms ties with different communities.

Another example can be given using a pollination network. In this case, the nodes are plants and pollinators such as bees, birds and butterflies, and the edges represent interactions: which pollinator visits which plant. Most pollinators have relatively stable foraging behaviour, and they prefer the same set of plants. But a roaming agent would be a generalist pollinator, like a bumblebee. On one day, it visits mostly wildflowers and on another, it shifts to orchards, or it moves to an entirely different set of plants. Such behaviour might depend on season, weather, or availability. Ecologically, this helps maintain biodiversity and ensures cross-pollination across communities.

\subsection{Local interaction group diversity}

The main aim of the model studied in this \docName is the ability to include the dynamics into the process of interaction neighbourhood selection. For this purpose, we defined an extension of the PGG model proposed in \cite{shang2022cooperation} enriched with the ability to reevaluate the assigned interaction neighbourhoods.

The pseudocode for the model, including the reevaluation of the interaction neighbourhood, is provided in \lstref{model:local-roaming}. We assume that there is exactly one agent located in each node of the 2D lattice. Hence, as in the case of such models, the geometry of the lattice dictates the possible selection of the interaction neighbourhood and the agents are limited to choosing interacting partners from von Neumann or Moore neighbourhoods.

\begin{algorithm}[t]
\begin{algorithmic}
    \REQUIRE size of the grid $L$, participation of roaming agents $\delta \in[0,1]$, number of agents $n=L^2$
    \STATE{} \COMMENT{Initialization loop -- executed once}
    \FOR{$i=1$ to $n$}
        \STATE $G_i$ $\leftarrow$ sub-population from the spatial neighbourhood  of $i$ \COMMENT{Assign interaction neighbourhood}
        \STATE{$s_i$ $\leftarrow$ 0 (defector) or 1 (cooperator)} \COMMENT{Assign initial strategy}
        \STATE{$\delta_i$ $\leftarrow$ 0 (stationary) or 1 (roaming)} \COMMENT{Assign roaming status}
    \ENDFOR

    \STATE{} \COMMENT{Evolution loop -- executed during each step, consists of two phases}

    \STATE{} \COMMENT{Phase 1 -- cumulate payoffs by playing PGG}
    \FOR{$i=1$ to $n$}
        \STATE{play PGG with agents in $j \in G_i$} \COMMENT{Assign payoff for each agent in $G_i$}
        \STATE{update income $\Theta_j$ for all $j\in G_i$}  \COMMENT{Accumulate payoffs from all elementary games}
    \ENDFOR

    \STATE{} \COMMENT{Phase 2 -- updated strategy, optionally update the interaction neighbourhood}
    \STATE{} \COMMENT{This can be done synchronously or asynchronously}
\FOR{$i=1$ to $n$} 
        \STATE j $\leftarrow$ {one of the neighbours from the interaction neighbourhood of $i$}
       	\STATE $\Delta \leftarrow \Theta_i - \Theta_j$ \COMMENT{Calculate difference of incomes}
        \STATE {$s_i$ $\leftarrow$ new strategy selected using $\Delta$} \COMMENT{Update using the selected imitation process}
        \IF{$\delta_i = 1$} 
        \STATE $G_i$ $\leftarrow$ sub-population from the spatial neighbourhood  of $i$ \COMMENT{New interaction neighbourhood}
 \ENDIF
    \ENDFOR
\end{algorithmic}
\caption{Public Goods Game with roaming agents on 2D grid (local scenario).}
\label{model:local-roaming}
\end{algorithm}

One should note two elements of the presented model. 

First, the spatial neighbourhood, used in the process of interaction neighbourhood selection, is one of the parameters. In the rest of this \docName, we assume that in the local scenario the agent is using von Neumann or Moore neighbourhoods, as these are the two natural neighbourhoods, studied in the situation where agents are placed on the 2D lattice. However, one should note that this option was chosen mainly to make the presented results relevant for the cases studied in the literature.

Second, all agents use one type of the spatial neighbourhood during the game. Hence, the maximal number of interaction local neighbours to select from is fixed. Hence, roaming agents are limited in their choice during the evaluation of the interaction neighbourhood, as they are not flexible in their choices. This issue is partially addressed by the model enabling global spatial interaction diversity model discussed in the next subsection.

One should also note that by introducing a new characteristic of the agents participating in the PGG we effectively diversify the process regarding the characterization of the participants. Usually, models where the agents are diversified regarding their characterization are built upon the diversification of the relations with other agents 
\cite{kurzban2001individual,coats2005beliefs}, beliefs and social preference
\cite{lucas2014effects,polaniareyes2015identification} or expectations \cite{chengyi2012effects}, which might in turn affect the mobility. On the other hand, the mobility preferences considered in this \docName are the intrinsic characteristic of the agent. This makes the presented model suitable to study any situation where the agents can alter the pool of agents they are interact with.

\subsection{Global interaction group diversity}

The geometry of the interaction dictated by the 2D lattice is the most important   factor limiting the interaction between the agents in the model presented in \lstref{model:local-roaming}. Hence, following the considerations from \cite{santos2008social}, it is natural to allow the formation of the interactions neighbourhoods by using a more realistic graph of connections.

To take this into account, we consider a scenario where the interaction neighbour is selected from the full population. We restrict only the number $K$ of interacting agents included in the interaction neighbourhood. 

The pseudocode for the global variant of the model with global interaction neighbourhoods is  included in~\lstref{model:global-roaming}. Again, as in the local case, we assume that there is exactly one agent assigned to each position. However, in the global case the spatial position of the agent is not relevant for the selection of the interaction neighbourhood as the spatial neighbourhood, representing the possible network of interactions, is given by a complete graph.

In general, in the global scenario, each agent is assigned a random number of interacting neighbour $ 1 \leq k_i\leq K$, with uniform distribution. However, one can also consider a version of the global scenario with the fixed number of the interacting neighbours assigned to each agent. In such case, the size of the interaction neighbourhood is not randomly chosen from $\{1,2,\dots,K\}$, but it is equal to $K$ for all agents, and hence $k_i=K$ for all  $i=1,2,\dots,L^2$. On the other hand, the number of interacting neighbours does not necessary have to be uniformly distributed. This situation lead to the different model of the random graph formed by the connections between the interacting neighbours~\cite{santos2008social}.

\begin{algorithm}[H]
\begin{algorithmic}
    \REQUIRE size of the grid $L$, participation of roaming agents $\delta \in[0,1]$, number of agents $n=L^2$, maximum number of interacting agents $K$
    \STATE{} \COMMENT{Initialization loop -- executed once}
    \FOR{$i=1$ to $n$}
        \STATE{$k_i$ $\leftarrow$ $U(\{1,2,\ldots,K\}$) } \COMMENT{Assign the size of the interaction neighbourhood}
        \STATE{$G_i$ $\leftarrow$ sub-population of $k_i$ agents} \COMMENT{Assign interaction neighbourhood}
        \STATE{$s_i$ $\leftarrow$ 0 (defector) or 1 (cooperator)} \COMMENT{Assign initial strategy}
        \STATE{$\delta_i$ $\leftarrow$ 0 (stationary) or 1 (roaming)} \COMMENT{Assign roaming status}
    \ENDFOR

    \STATE{} \COMMENT{Evolution loop -- executed during each step, cosists of two phases}

    \STATE{} \COMMENT{Phase 1 -- cumulate payoffs by playing PGG}
    \FOR{$i=1$ to $n$}
        \STATE{play PGG with agents in $j \in G_i$} \COMMENT{Assign payoff for each agent in $G_i$}
        \STATE{update income $\Theta_j$ for all $j\in G_i$}  \COMMENT{Accumulate payoffs from all elementary games}
    \ENDFOR

    \STATE{} \COMMENT{Phase 2 -- updated strategy, optionally update the interaction neighbourhood}
     \STATE{} \COMMENT{This can be done synchronously or asynchronously}
\FOR{$i=1$ to $n$} 
        \STATE j $\leftarrow$ {one of the neighbours from the interaction neighbourhood of $i$}
       	\STATE $\Delta \leftarrow \Theta_i - \Theta_j$ \COMMENT{Calculate income difference}
        \STATE {$s_i$ $\leftarrow$ new strategy selected using $\Delta$} \COMMENT{Update the strategy using the selected imitation process}
        \IF{$\delta_i = 1$} 
        \STATE{$k_i$ $\leftarrow$ $U(\{1,2,\ldots,K\}$) } \COMMENT{Assign new size of the interaction neighbourhood}
        \STATE $G_i$ $\leftarrow$ sub-population from the spatial neighbourhood  of $i$ \COMMENT{New interaction neighbourhood}
 \ENDIF
    \ENDFOR
\end{algorithmic}

\caption{Public Goods Game with roaming agents in the global scenario.}
\label{model:global-roaming}
\end{algorithm}

\section{Results}\label{sec:results}

In this section, we will present numerical results illustrating the impact of the dynamical elements in the interaction group diversity on the Public Goods Games. The implementation of based on the local and the global versions of the presented model and the source code can be found at \cite{pgg-spatial-model-mc}.

We focus on two aspects of the dynamics. 

The first aspect is related to the process of neighbourhood selection. In our case, in the local regime, the interaction neighbourhood can be selected as a fixed size group of agents forming von Neumann neighbourhood or Moor neighbourhood of the agent. Following~\cite{shang2022cooperation}, we also consider the case where the agent can select only some of the neighbours from the spatial neighbour to be included in the interaction neighbourhood.
On the other hand, in the global regime, the interaction neighbourhood is selected as a fixed size neighbourhood with $K$ agents from the full population.

We argue that the inhomogeneity in the process of neighbourhood selection leads to the lower values of the synergy factor $r$ required to achieve the collaboration among the agents, not only in the local scenario, but also in the case of using global interactions.

The second aspect is used to introduce the possibility of reevaluation of the interaction neighbourhood. This is achieved by introducing the subpopulation of \emph{roaming} agents, which can alter the selected interaction neighbourhood. For the sake of simplicity, we fix a probability of reevaluation, and the additional parameter in the model is the roaming agent participation $\delta$.

We argue that by introducing a limited subpopulation of roaming agents, one can decrease the synergy factor required to achieve a domination of cooperators in the population. Hence, one can conclude that the diversification of the interaction neighbourhood, understood as the inhomogeneity in the interaction neighbourhood and the ability to alter it, leads to a better cooperation.

Following the setup from \cite{shang2022cooperation}, we set the thermal factor in Eq.~(\ref{eqn:imitation-fermi}) to $\kappa=0.5$. Thus, the impact of the noise is constant and does not contribute to the changes in the formation of cooperation. Moreover, our figure of merit is the average number of cooperators in the last $2^{10}$ steps of the evolution. This will ensure that the obtained results are stable and do not depend on the short-term fluctuations.

In \cite{shang2022cooperation}, to consider the inhomogeneity of the interaction neighbourhoods, the situation with players having different number of interacting agent has been considered. The major difference of the presented results is that they take into account the possibility of altering the interaction neighbourhood. This, along with the lack of restrains imposed by the 2D lattice geometry, makes the presented model more flexible for the purpose of simulating behaviour observed in realistic populations.

Also, one should note that the main figure of merit in this \docName is the average number of cooperators in the last $10^3$ steps. This quantity is visualized in all plots in this section.

\begin{figure}
	\begin{center}
		\includegraphics[width=0.4\textwidth]{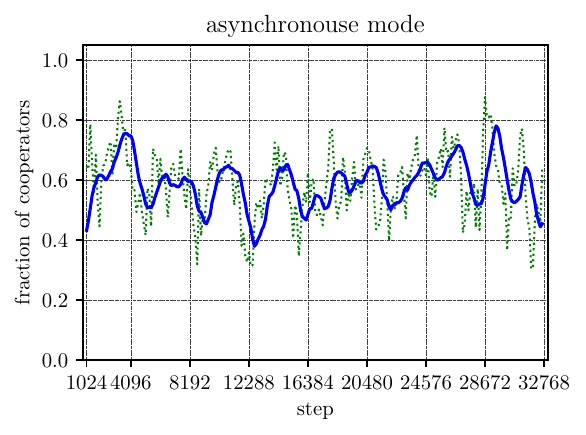}\hspace{2em} \includegraphics[width=0.4\textwidth]{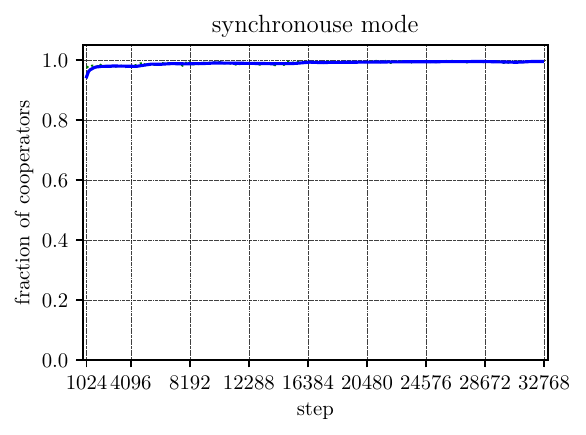}\\
		\caption{Examples of the limiting behaviour in asynchronous (left) and synchronous (right) realizations. The green dotted line illustrates the number of cooperators, the blue solid line illustrates the average number of cooperators in the last $10^3$ steps, which is the figure of merit in \cite{shang2022cooperation} and in this manuscript. In this case we have: grid size $L\times L = 64\times64$, synergy factor $r=6$, random $K$ patches with $K=6$, roaming agents fraction $\delta=0.1$, and $2^{15}$ time steps.}
		\label{fig:examples-limiting}
	\end{center}
\end{figure}

An example realization of the considered process for variants with the synchronous and asynchronous strategy update are presented in Fig.~\ref{fig:examples-limiting}. As one can see from this example, the updated mode has a significant impact on the stability of the system, and one can expect that the synchronous mode will lead to the more prominent impact of agent's mobility on the system behaviour.

\subsection{Local scenario}

Let us start by considering the case where the agent is limited to choose its interaction partner from von Neumann or Moor neighbourhood. In the model introduced in \cite{shang2022cooperation}, each agent is assigned an interaction neighbourhood, which consists of a subset of its von Neumann or Moore neighbourhood. Assigned interaction neighbourhoods do not change during the evolution.

In the situation considered in this \docName, in the initialization step each agent is assigned interaction neighbourhoods, chosen from von Neumann neighbourhood or Moore neighbourhood. This corresponds to \emph{random von Neumann} and \emph{random Moore} variants. Additionally, we include the roaming agents, which can change their interaction neighbourhood with probability $\frac{1}{2}$ after each round.

\begin{figure*}[ht!]
\centering
\subfigure[imitation using Fermi function (Eq.~\ref{eqn:imitation-fermi}), synchronous update]{\includegraphics[]{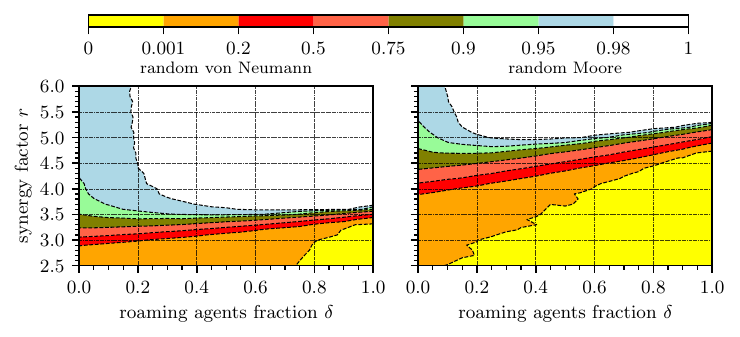}\label{fig:random-local-roaming-fd}}

\subfigure[imitation based on the differences of payoffs (Eq.~\ref{eqn:imitation-diff}), asynchronous update]{\includegraphics[]{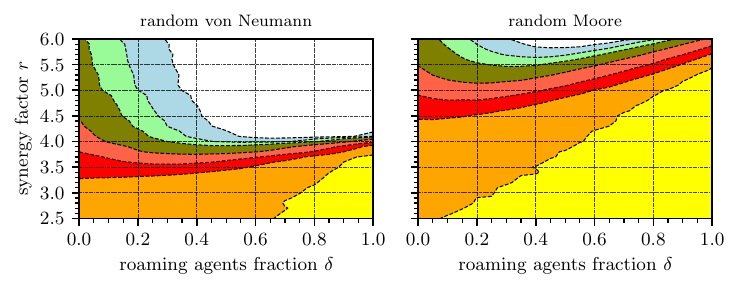}
\label{fig:random-local-roaming-diff}
}

\subfigure[imitation using Fermi function  (Eq.~\ref{eqn:imitation-fermi}), asynchronous update]{\includegraphics[]{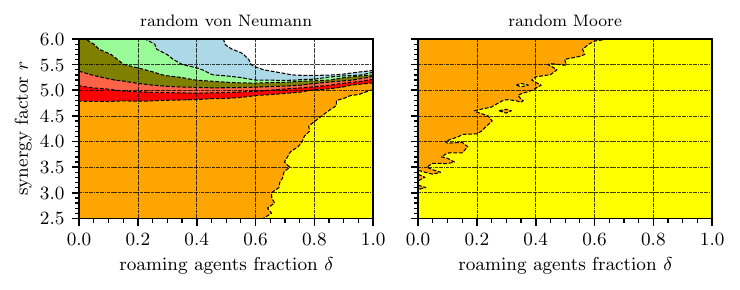}
	\label{fig:random-local-roaming-fd-ascync}
}

\caption{Impact of roaming agents on the formation of cooperation in the local scenario. The interaction group is selected from von Neumann (left panels) and Moore (right panels) neighbourhoods. Plots represent the average number of cooperators after $2^{10}$ steps, on $d=L\times L = 64\times64$ grid with periodic boundary conditions. The case $\delta=0$ in \subref{fig:random-local-roaming-fd} corresponds to the results obtained in \cite{shang2022cooperation} for the situation where the interaction neighbourhoods are random and fixed. In the case \subref{fig:random-local-roaming-diff} the participation of roaming agents is even more important than in the case of the Fermi imitation function, and it is necessary to increase the fraction of cooperators in the population. One can also note that in the asynchronous update mode \subref{fig:random-local-roaming-fd-ascync} the effect of the roaming agents is still present, albeit it is less visible and almost vanishes for the Moore neighbourhood in the considered range of parameters. }
\label{fig:random-local-roaming}
\end{figure*}

In the model considered in this \docName, a subpopulation of roaming agents can reevaluate the assigned interaction neighbourhood. Hence, for the roaming agents participation $\delta=0$, we reproduce the model and the results from~\cite{shang2022cooperation}. A similar effect was observed in~\cite{hu2021unfixed}, where it was numerically demonstrated that by altering the number of interactive neighbours  in each round one can boost the cooperative behaviour. As in the case of Fig.~\ref{fig:random-local-roaming-diff}, the results obtained in~\cite{hu2021unfixed} were confirmed not only for the Fermi function. For this reason, one can argue that the observed effect is not related to the specific updated rule. Hence, in the rest of this \docName we focus on the numerical results obtained using Fermi function only as it is has been used in \cite{shang2022cooperation}, where the initial version of the discussed model was considered.

One should also note that the advantage of the presented model in comparison to the unfixed-neighbour mechanism introduced in~\cite{hu2021unfixed} is that it can be used to gradually control the dynamics in the neighbourhood selections. This allows us to study how the changes in the level of collaboration are affected by the changes in the interaction neighbourhoods.

This can be observed in Fig.~\ref{fig:random-local-roaming} where the impact of the roaming agents participation $0\leq\delta\leq1$ on the collaboration level is presented. As one can note, for $\delta=0$, the cooperation strategy is dominating in the population for the synergy factor $r_{vN}>3.5$ in the case of von Neumann neighbourhood and for $r_{M}>4.75$ in the case of Moore neighbourhood (cf. Fig.~1 in \cite{shang2022cooperation}).

One should note that the negative impact of the heterogeneity on the cooperation was also observed in previous studies.
In particular, the effect is similar to the effect observed in~\cite{flores2023heterogeneous}, where there exists contribution optimal value of the contribution for minimizing the synergy factor required to form the cooperation.

\subsection{Global scenario}

In the second series of the experiments, we investigate the impact of the roaming agents participation on the PGG in the case where the global scenario for spatial interaction diversity is considered. This corresponds to the model described in Listing~\ref{model:global-roaming}. In this case, the agents are not restricted to interact only with the agents in their vicinity; hence, there are no restrictions resulting from the 2D lattice geometry. We can distinguish two basic variants of the situation -- the scenario with fixed $K$, where we introduce only the freedom in group diversification, and the scenario with fixed maximal $K$, where we introduced the freedom of selecting the group and the group size.

\subsubsection{Diversification of the group}

\begin{figure*}[ht!]
\centering
\includegraphics{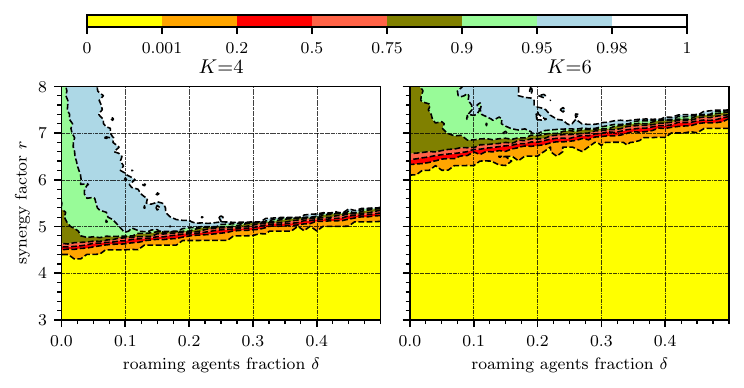}
\caption{Impact of the roaming agents in the global scenario with the fixed number of interacting agent $K$. In this case, the effect of boosting cooperation by introducing roaming agents is clearly visible. However, for the larger values of $K$ the system is struggling to achieve a high rate of collaboration. Results presented in the figure were obtained for the model with $d=L\times L=32\times 32$ and periodic boundary conditions, averaged over 50 realizations, each consisting of $2^{11}$ steps, synchronous update mode.}
\label{fig:plot_roaming-patches-l32-even}
\end{figure*}

The simplest extension of the situation discussed above is given by the scenario where the agents are free to choose their interaction group from the full population. This means that the interaction group of each agent is defined by the full graph of connections between agents.

Let us first focus in the case where the freedom of choosing the interaction group is limited by fixing the number of interacting agents $K$ only. This corresponds to lifting the spatial restriction of the 2D lattice geometry, which fixes the number of interacting agents. In the case of the local scenario with von Neumann or Moore spatial neighbourhoods, this would be equivalent to a static model with fixed interaction group. This effectively changes the geometry of the interactions into the complete graph.

In this situation, each agent can interact not only with the agents it will select, but also with the agents selecting it. Indeed, the average number of interacting neighbours is given by
\begin{equation}
\langle N_i^K \rangle = K \left(1 + \frac{d-(K-1)}{d-1}\right),
\end{equation}
where $d$ is the size of the graph, which is in this case $d=L\times L$. With the increase sized in the of the lattice, we get the scaling
\begin{equation}
\langle N_i \rangle \approx 2 K.
\label{eq:ni-scale-maxk}
\end{equation}

The relation between the presence of roaming agents and the cooperation formation is in the global scenario with fixed number of interaction group is plotted in Fig.~\ref{fig:plot_roaming-patches-l32-even}.
The presented results confirm that the quantitative behaviour of the introduced model is similar to the situation considered in the local scenario. The major difference in numerical values obtained in the local and in the global scenarios result from the impact of the freedom in the interaction selection on the effective size of the interaction groups. One should note that considering Eq.~\ref{eq:ni-scale-maxk}, the situation with $K=4$ in the global scenario (cf. left panel in Fig.~\ref{fig:plot_roaming-patches-l32-even}) should be compared with the local scenario where the interaction neighbourhoods are selected using the Moore spatial neighbourhood.

Additionally, it is visible that the lack of the restricted interaction geometry results in the decrease in the range of process parameters for which the system does not converge either to the state of full cooperation nor to the state of full defection. Hence, one can argue that the lack of constrains in the neighbourhood selection, and the resulting increase in the effective number of the interacting neighbours, leads to a less stable behaviour than in the case of local interaction neighbourhoods. 

\subsubsection{Diversification of group and size}

\begin{figure*}[ht!]
\centering
\includegraphics[]{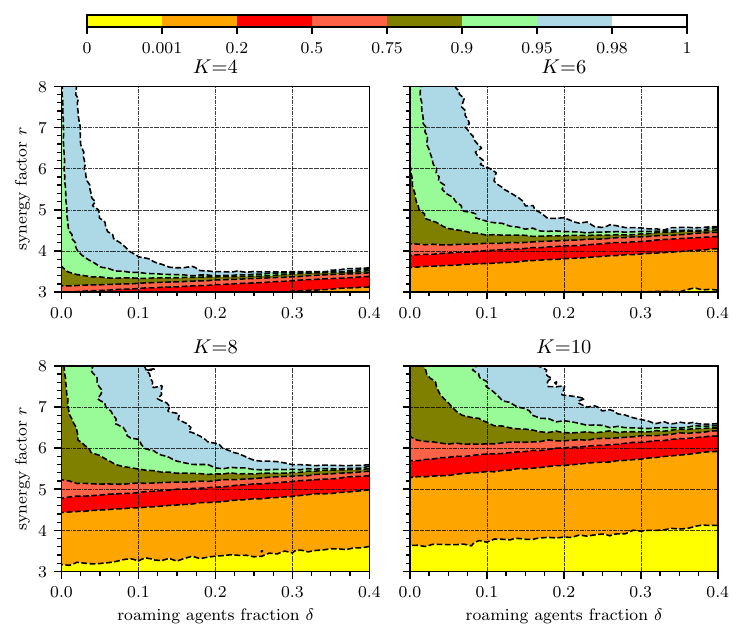}
\caption{Impact of the increasing participation of roaming agents on the formation of cooperative behaviour in the case of random interaction groups. The results represent values of the average in the last $2^{10}$ steps, averaged over 50 realizations, for $K=4,6,8,10$ and for $0\leq\delta\leq0.4$. Plots were obtained for square $L\times L$ grid, with $L=64$ and periodic boundary conditions. Each configuration was evaluated for $2^{11}$ steps. Average value of cooperators in the last $10^3$ steps was calculated over $50$ realizations, synchronous update mode.}
\label{fig:random-patches-roaming}
\end{figure*}

In the global scenario, each agent can select its interaction neighbourhood from among the full population. Each agent can select up to $K$ interacting neighbours. In our numerical experiments we consider the cases where $K=3,4,\dots,10$. One should also note that no particular structure on the distribution of the number of interacting partners is assumed. Hence, when considering a case with parameter $K$, the agent can choose uniformly from the set $\{1,2,\dots, K\}$. Thus, the obtained graph of connections does not have a scale-free structure.

In this case, considering that the average number of the chosen group is $\frac{1+K}{2}$, we get that the average size of the interaction group will be given by
\begin{equation}
\langle N_i^{U(K)} \rangle = \frac{1+K}{2} \left(1 + \frac{d-(K-1)}{2(d-1)}\right),
\end{equation}
Hence, for the size of the lattice $d\mapsto \infty$ we get
\begin{equation}
\langle N_i^{U(K)} \rangle \approx 1 + K,
\end{equation}
resulting in the average size of the interaction neighbourhood significantly smaller than in the case of the global scenario with fixed $K$.

From the plots presented in Fig.~\ref{fig:random-patches-roaming} one can observe that for $K>3$ some participation of roaming agents can be beneficial for cooperation. Indeed, for the roaming agent's participation $\delta$ in the range $\delta\in[0.1,0.5]$ it is visible that the roaming agents decrease the synergy factor $r$ required to achieve the domination of the cooperative strategy.

As one can see in the plots presented in Fig.~\ref{fig:random-patches-roaming}, in the case where the interaction neighbourhood can be selected from the global pool of agents with the uniformly distributed size of the interaction neighbourhood, one can also observe the effect of decreasing of the synergy factor 

Moreover, from the results presented in Fig.~\ref{fig:random-patches-roaming} one can see that, in comparison to the global scenario with the fixed $K$, the case with the diversification in the interacting neighbourhood size leads to more stable behaviour of the system. This allows us, as in the case of the local scenario, to distinguish the minimal value of the roaming agent's participation which allows achieving the state of full cooperation.

\subsection{Scaling of the roaming impact}
Using the data presented in Figs.~\ref{fig:random-local-roaming} - \ref{fig:random-patches-roaming}, one can notice that the synergy factor required to achieve a given level of cooperation depends on the fraction of roaming agents, $\delta$. Hence, one can consider the synergy factor as a function of $\delta$, namely $r = r(\delta)$. 

\begin{figure*}[ht!]
\centering
\includegraphics{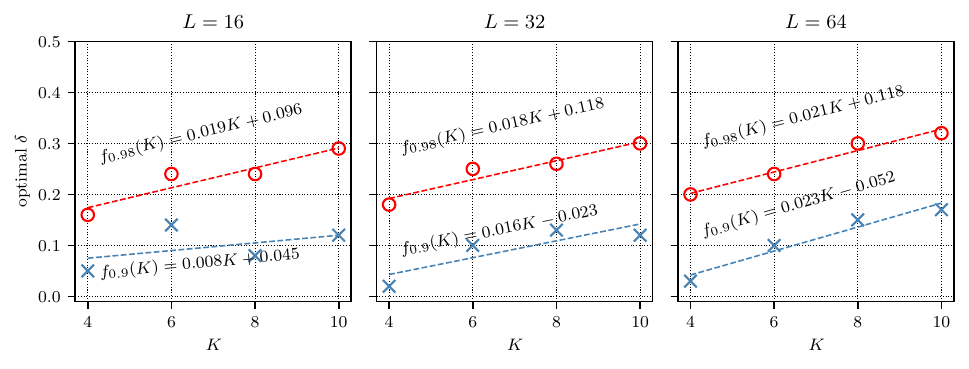}
\caption{Optimal participation of roaming agents minimizing the synergy factor required for achieving the cooperation with $0.90$ (blue crosses) and $0.98$ (red circles) for increasing values of maximal interaction neighbourhood size $K$. In the left plot, the data were obtained for a case with $L=16$ using the averaging over 150 realizations. In the middle plot the data were obtained using $L=32$  and in the right plot data were obtained for $L=32$. Data for $L=32$ and $L=64$ were obtained from averaging over 50 realizations.}
\label{fig:min-delta}
\end{figure*}

Let us denote by $\mathcal{C}(PGG(\delta,r))$ is the mean fraction of cooperators, achieved in the Public Goods Game with synergy factor $r$ and participation  of roaming agents $\delta$.

The most interesting observation from the performed numerical experiments is that the participation of the roaming agents is reaching the optimal point in each one. At this point, the synergy factor which is required to ensure that the cooperation is dominant becomes minimal, and it cannot be decreased without increasing the synergy factor required to achieve the threshold $T_\mathcal{C}$.

From the obtained results one can argue that there is an optimal value of the roaming agent's participation, $\delta^\mathcal{C}$, for each threshold of the cooperator's participation. At this value, the system is achieving the ratio of collaborators at the level of $T_\mathcal{C}$, with the minimal synergy factor required to achieve this. This is equivalent to saying that the curve $\{(\delta,r(\delta)) : \mathcal{C}(PGG(\delta,r)) = T_\mathcal{C}\}$ is the Pareto front of the problem of joint optimization of the cooperation participation regarding the roaming agents participation and the synergy fraction. Hence, $\delta^\star$ is defined as
\begin{equation}
\delta^\star = \min_\delta \min_r \{r(\delta): \mathcal{C}(PGG(\delta,r)) \geq T_\mathcal{C}\} 
\end{equation}

In the particular examples presented in this \docName we fix the threshold of the cooperator's participation $T_\mathcal{C}=0.98$. Additionally, by the cooperator's participation, we understood the average number of cooperators in the last $2^{10}$ Monte Carlo simulation steps.

The plots presented in Fig.~\ref{fig:min-delta} demonstrate the scaling of the optimal $\delta^\star$ for the cases of Public Goods Game with roaming agents on lattices with the interaction neighbourhoods of size $4\leq K \leq 12$, with varying size of the lattice set to $L=16$, $L=32$, and $L=64$. One can see that the scaling of the optimal value of the roaming participation is very similar and is independent of the size of the lattice. This demonstrates that the introduction of the roaming agents should not be governed by the size of the considered system, but rather by the capabilities of the agents to choose their interaction neighbourhoods.

\subsection{Comparison with the base model}

Let us now discuss the difference between the introduced model and the base model proposed in~\cite{shang2022cooperation}.

\begin{figure*}[ht!]
	\centering
\includegraphics{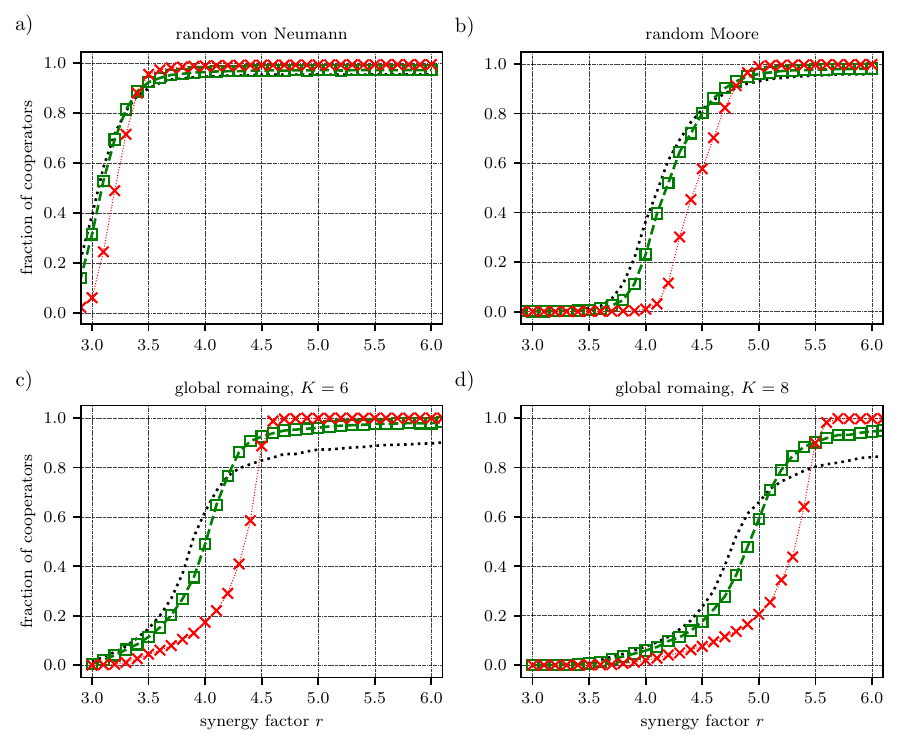}
\caption{Comparison of the static model from \cite{shang2022cooperation} and the model with the roaming agents in the local and the global variant. The plots were prepared for $\delta=0$ (black dotted line), $\delta=0.1$ (green squares), and $\delta=0.4$ (red crosses). In two top plots and for the case  $\delta=0$, the base model from \cite{shang2022cooperation} is obtained.
	Plots were obtained for square $L\times L$ grid, with $L=64$ and periodic boundary conditions. Each configuration was evaluated for $2^{11}$ steps.
	Average value of cooperators at the final step, after $2^{10}$ steps, calculated over $50$ realizations, synchronous update mode.}
\label{fig:roaming-vs-static}
\end{figure*}

Firs of all, if we restrict the ability of agents to chose randomly from von Neumann or Moor neighbourhoods only once, the presented model is reduced to the cases discussed in~\cite{shang2022cooperation}. This is visible in Fig.~\ref{fig:roaming-vs-static}, where the top plots illustrate the situation with the local interaction neighbourhoods. In this case, the interaction neighbourhoods are inhomogeneous due to the randomization of the interaction partners in the initialization phase.

In particular, panels a) and b) in Fig.~\ref{fig:roaming-vs-static}, demonstrate plots and for the case  $\delta=0$, e.g. the case  where the agents are not roaming and the random selection of the interaction partners is done only once. One can observe in this situation that the scenario with the von Neumann neighbourhood lead to the faster formation of cooperation comparing to the scenario with the Moor neighbourhood (cf. Fig.~1 in~\cite{shang2022cooperation}).

Moreover, in this case, the model introduced in this \docName has a negligible impact on the formation of cooperation or can lead to the slower increase in  the number of cooperators if the participation of the roaming agents is significant. This is visible in Fig.~\ref{fig:roaming-vs-static} b), where for $\delta=0.4$, the participation of cooperators grows slower with the increasing synergy factor.

On the other hand, considering the global version of the model presented in this \docName, one can see that the moderate participation of the roaming agent can be beneficial for the cooperation. Again, for the global case with $K=6$, which can be seen as a generalization of the scenario based on the von Neumann neighbourhood, the formation of cooperation is easier when compared to the case $K=8$.

In Fig.~\ref{fig:roaming-vs-static} c) and Fig.~\ref{fig:roaming-vs-static} d), the participation of the roaming agents on the level of $\delta=0.1$ leads to a higher level of cooperating agents for the considered values of the synergy factor. However, by increasing the participation of roaming agents to  $\delta=0.4$, the increase in the number of cooperating agents is slower, even comparing to the case where there are not roaming agents in the population at all.

\section{Conclusions and final remarks}\label{sec:final}

The goal of the presented \docName is to study the impact of agents changing their interaction neighbourhoods on the formation of collaboration. Motivated by the models described in \cite{santos2008social} and \cite{shang2022cooperation} we have provided a model capturing the connection between the cooperation and the diversity in the interaction groups.

We have extended a model of spatial interaction diversity proposed in~\cite{shang2022cooperation} by enabling the reevaluation of the interaction neighbourhoods by a subpopulation of roaming agents. We also studied the model of interaction diversity limited to a fixed, local neighbourhood, as well as in the case where the agent is not restricted by the geometry of the spatial graph.

In contrast to \cite{santos2008social}, we observe the effect of boosted cooperation in a simple model where the subpopulation of agents is allowed to reevaluate their interaction neighbourhood. 
The model proposed in \cite{santos2008social} assumed that the structure of interaction neighbourhoods follows the structure of scale-free network. Hence, the results presented in this \docName suggest that by tuning the participation of roaming agents in the population, one can mimic the behaviour observed in the scale-free networks. However, the presence of roaming agents can also be destructive for collaboration, and goes beyond the model from~\cite{santos2008social}.

Moreover, in contrast to the models introduced in \cite{hu2021unfixed}, the model proposed in the presented \docName can be used to relate the level of dynamics in the neighbourhood selection to the formation of cooperative behaviour. Using the presented model, one can control the level of dynamics by controlling the number of roaming agents. Hence, the presented model can be used to emulate the unfixing of the interaction neighbourhood introduce in~\cite{hu2021unfixed}.

One should also note that the parameter $K$ considered in the \docName is used to generalize the number of agents in the Moore of von Neumann neighbourhood, which is commonly used in the studies involving PGG. In particular, the aim of this paper was to generalize the model from \cite{shang2022cooperation}, where Moore and von Neumann neighbourhoods are used. In the presented \docName values up to $K=10$ are considered, with only values up to $K=8$ corresponding to Moore and von Neumann cases. However, the model was not tested with the group size $K$ as the main parameters, and the main figure of merit was the fraction of roaming agents in the population.

We observe how the optimal roaming agents participation, minimizing the required synergy factor, grows with the increase in the size of the interaction group. The obtained results suggest that the significant degree of the synergy can be achieved even for a small fraction of roaming agents present in the population. On the other hand, we also observed that the unlimited growth of the number of roaming agents has a negative result on the formation of cooperation.

The reevaluation of the interaction neighbourhood considered in the presented \docName can be interpreted as the engagement in various social interactions. Hence, the results of the presented \docName confirm that interacting with different groups within the population facilitates the formation of cooperation. For this reason, one can conclude that the diversification of the interaction neighbourhood, understood as the ability to alter it and being able to interact with new partners, leads to improved cooperation between the agents.

However, one should note that the presented results are based on the numerical evaluation of the simplified model. As such, they be also understood as a hint in the process of understanding the social complexity~\cite{anzola2022ethics}.

\paragraph{Acknowledgements}
Author would like to thank Izabela Miszczak for interesting discussions concerning the management of the commons and constructive comments received from the anonymous reviewers. Moreover, the author would like to acknowledge that his work has been motivated by personal curiosity and
received no support from any agency.

We gratefully acknowledge Polish high-performance computing infrastructure PLGrid (HPC Center: ACK Cyfronet AGH) for providing computer facilities and support within computational grant no. PLG/2023/016068.
\paragraph{Code availability}
The implementation of the presented model can be found at \cite{pgg-spatial-model-mc} and \cite{miszczak2025pgg-group-diversity}. Additionally, \cite{miszczak2025pgg-group-diversity} contains definition of experiments, scripts used to control numerical experiments, scripts for plotting results, as well as data used in the presented plots.



\end{document}